\documentclass[conference]{IEEEtran}
\IEEEoverridecommandlockouts
\usepackage{verbatim}%for the comment
\usepackage[utf8x]{inputenc}
\usepackage{epstopdf}
\usepackage[pdftex]{graphicx}
\usepackage{tabularx}
\usepackage{booktabs}
\usepackage[table]{xcolor}
\usepackage{booktabs}
\usepackage{multirow}
\usepackage{ctable}
\usepackage[cmex10]{amsmath}
\usepackage{algorithm}
\usepackage{algpseudocode}
\usepackage{pifont}
\usepackage{amssymb}
\usepackage{cite}
\usepackage{textcomp}
\usepackage{array}
\usepackage{amsthm}
\usepackage{wrapfig}
\usepackage{subcaption}
\graphicspath{{./Figures/}}

\theoremstyle{plain}
\newtheorem{lemma}{Lemma}
\newtheorem{theorem}{Theorem}

\newcommand{\g}{\operatorname{g}}
\newcommand{\air}{\operatorname{a}}
\newcommand{\CI}{\operatorname{CI}}
\newcommand{\CN}{\operatorname{CN}}

\begin{document}
\bstctlcite{IEEEexample:BSTcontrol}
%---------------------------------------------------------------------------------
%                                    Title
%---------------------------------------------------------------------------------
\title{Performance Analysis of Uplink Cellular IoT Using Different Deployments of Data Aggregators} %The title of the paper
\author{\IEEEauthorblockN{Ghaith Hattab and  Danijela Cabric}
	\IEEEauthorblockA{Department of Electrical and Computer Engineering\\
		University of California, Los Angeles\\
		Email: ghattab@ucla.edu, danijela@ee.ucla.edu}
\thanks{This work has been supported by the National Science Foundation under grant 1149981.}
}

\maketitle

%---------------------------------------------------------------------------------
%                                    Abstract
%---------------------------------------------------------------------------------
\begin{abstract}
Data aggregation is an effective solution to enable cellular support of Internet-of-things (IoT) communications. Indeed, it helps alleviate channel congestion, reduce the communication range, and extend battery-lifetime. In this paper, we use stochastic geometry to analyze the performance of uplink cellular IoT using different deployment strategies of aggregators, including terrestrial and aerial ones, e.g., drones or unmanned aerial vehicles. We focus on IoT-specific performance metrics, that are typically used by 3GPP.  Specifically, we derive closed-form expressions of the average transmit power consumption, which is key to determine the lifetime of IoT devices, as well as the maximum coupling loss, which is essential to determine the maximum coverage the cellular system can support. Simulation results are presented to validate the derived theoretical expressions. It is shown that aerial aggregators can significantly extend the device lifetime and provide superior coverage compared to other deployment strategies. In addition, random deployment performs well when aggregators are densely deployed, whereas optimizing the location of a single terrestrial aggregator is beneficial when devices are more clustered.
\end{abstract}

%---------------------------------------------------------------------------------
%                                    Index Words
%---------------------------------------------------------------------------------

%---------------------------------------------------------------------------------
%                             I. Introduction
%---------------------------------------------------------------------------------
\section{Introduction}
Cellular Internet-of-things (IoT) is envisioned to bring a myriad of transformative opportunities for mobile network operators across various fields. Nevertheless, the support of IoT applications requires solutions tailored to their unique requirements compared to human-type traffic. For instance, many of these applications rely on the deployment of a large number of low-cost battery-powered IoT devices, with each device generating low-rate sporadic traffic \cite{DawyYaacoub2017}. In addition, some IoT objects are located deep indoors, e.g., utility meters. For these reasons, recent enhancements of cellular releases have recognized energy-efficiency and coverage as key pillars toward realizing large-scale IoT communications \cite{ShiraziBennett2015,Rico-AlvarinoYavuz2016}.

One approach to help support cellular IoT is the deployment of data aggregators, which act as relays, connecting IoT devices to the cellular network \cite{DawyYaacoub2017}. Indeed, by deploying aggregators in a given area, the coverage can be improved and the transmit power can be reduced as the communication distance decreases. They also help in signal congestion control, as a large number of IoT access requests can be condensed into fewer links with the core network. One key question is how such aggregators should be deployed in a given region. In this paper, we aim to study different deployment strategies of data aggregators using stochastic geometry. Specifically, we compare them in terms of (i) the average power consumption, which affects the lifetime of IoT devices, and (ii) the maximum coupling loss (MCL), which determines the coverage performance that can be supported by the network \cite{CATM2017}. 

Several works have studied the deployment and optimization of data aggregators via stochastic geometry. In \cite{KwonCioffi2013}, the optimal density of randomly deployed aggregators that meets an outage constraint of randomly deployed devices is given, in the absence of power control. In \cite{MalakAndrews2016}, the energy-efficiency of single-hop and multi-hop data aggregation is studied  in the presence of full-channel inversion, where the locations of aggregators and devices follow the homogeneous Poisson point process (HPPP). In this paper, we consider uplink (UL) fractional power control, study different deployment strategies besides the random deployment of aggregators, and assume IoT devices are clustered in a circular disk. In \cite{Guo2017}, the coverage performance and channel utilization of IoT devices are analyzed, where devices are centered around data aggregators. We note all aforementioned works consider terrestrial aggregators, whereas in this work we also study using aerial ones, e.g., unmanned aerial vehicles (UAVs) and drones. For instance, in \cite{Hattab2017a}, using drones as mobile data aggregators is considered from the perspective of the coexistence of IoT devices and legacy cellular users. In this paper, we instead focus on power consumption (or lifetime) and the MCL.

To summarize, the key contributions of this paper are as follows. We present different deployment strategies of data aggregators, highlighting their use cases. These deployments include (i) randomly deployed aggregators or small cells, (ii) randomly selecting a single or multiple devices as cluster heads, and (iii) optimizing the location of the aggregator to be at the centroid of the cluster, where the aggregator can be aerial or terrestrial. We then compare these schemes using stochastic geometry, focusing on key 3GPP metrics related to cellular IoT. Specifically, we find the average transmit power consumption, which is used then to determine the lifetime of IoT devices, following the 3GPP evaluation methodology \cite{3GPP2015e}. We also study the MCL, also denoted as the \emph{IoT coverage}, which is a metric used by 3GPP to evaluate the coverage performance of IoT-based solutions \cite{3GPP2015a}. The derived expressions help glean design insights on the deployment of data aggregators, and they are validated via Monte Carlo simulations. 

The rest of the paper is organized as follows. The system model and the deployment strategies are introduced in Section \ref{sec:model}. The performance analysis of the different schemes is presented in Section \ref{sec:analysis}. Simulation results are provided in Section \ref{sec:sim}, whereas the conclusions are drawn in Section \ref{sec:conclusion}.

%---------------------------------------------------------------------------------
%                II. System Model and Proposed Architecture
%---------------------------------------------------------------------------------
\section{System Model and Deployment Strategies}\label{sec:model}
We consider a cluster of devices that are uniformly distributed over an area of radius $R$. When $R$ is small, this amounts to sensors and machines having a similar task, e.g., deploying sensors in a farm for pest control, inside an infrastructure or a warehouse for monitoring, etc.  For very large values of $R$, this amounts to random deployment of sensors over large geographical areas. Each device is assumed to use  fractional power control for uplink (UL) data communications, as specified in \cite{3GPP2015a}. Specifically, assuming the device is at distance $r$ from the aggregator and the path loss is $l(r)$, then the transmit power is
\begin{equation}
\label{eq:TxPower}
P_{T} = \min\{P_o l(r)^{\epsilon\alpha},P_{\operatorname{max}}\},
\end{equation} 
where $P_o$ is the open-loop transmit power, $P_{\operatorname{max}}$ is the maximum allowable transmit power, $\alpha$ is the path-loss exponent, and $\epsilon\in[0,1]$ is the power control factor (PCF). Note that if an extended coverage mode is used, then $P_T= P_{\operatorname{max}}$ \cite{3GPP2015a}.
%
%hus, this power control algorithm requires knowledge of the path loss, which can be done by measuring the received signal and comparing it to the reference transmit power used in the downlink, as done in LTE networks \cite{3GPP2018d}. 

\begin{figure}[t!]
	\center
	\includegraphics[width=3.5in]{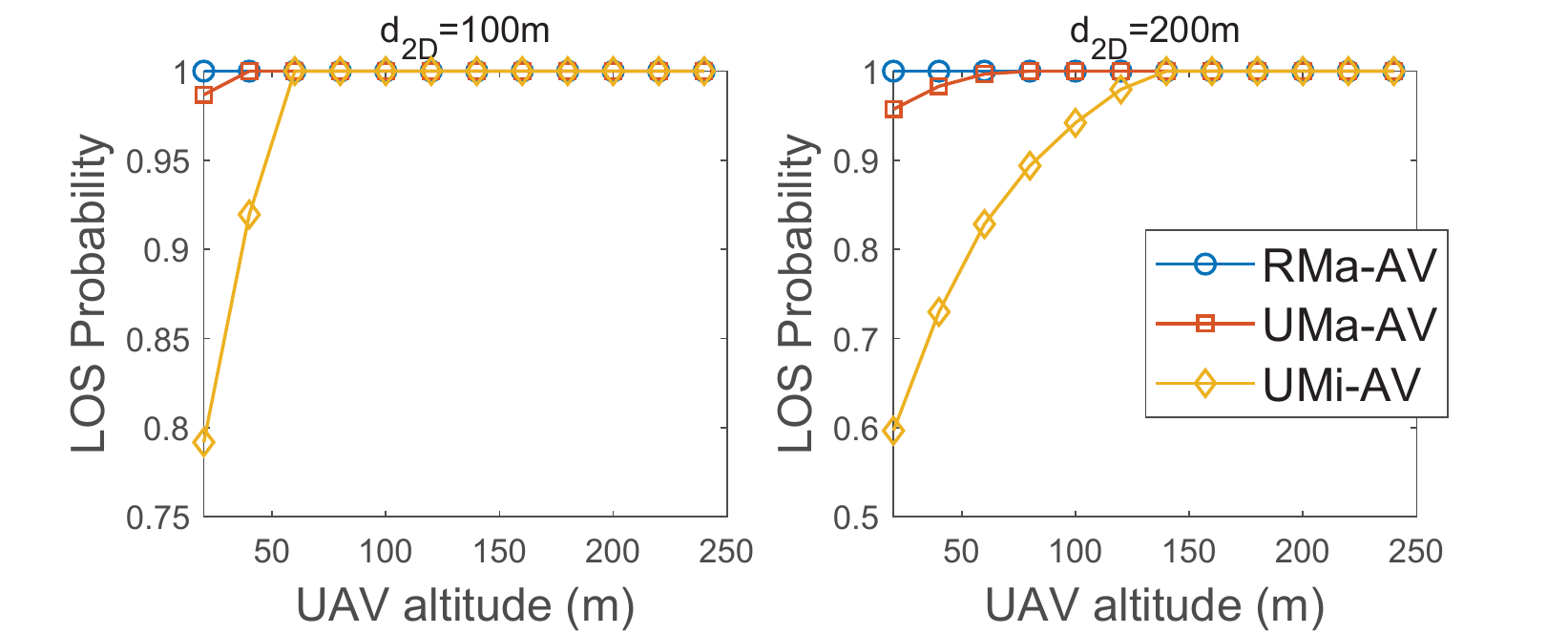} 
	\caption{LOS probability with variations of drone's altitude for different channel models.}
	\label{fig:PLOS_vs_height}
\end{figure}

For a ground-to-ground link between the device and a terrestrial aggregator, we consider the path loss $l_{\g}(r)=L_o r^{\alpha_{\g}}$, where $\alpha_{\g}$ is the path loss exponent and $L_o$ is the path loss at a reference distance of 1m. For ground-to-air links between IoT devices and an aerial aggregator, e.g., a UAV or a drone, it is shown in \cite{Al-HouraniLardner2014} that the channel can be decomposed into two groups: line-of-sight (LOS) and non-LOS (NLOS). In this paper, we assume that the aerial aggregator flies at an altitude such that the NLOS propagation occurs with very low probability, and hence it can be ignored as done in \cite{ZhangZhang2017,Ravi2016}.\footnote{Since we focus on the MCL, or the link budget, small-scale fading is ignored \cite{CATM2017}, yet it can be accounted for by adding a fade margin.} To this end, the path loss for the ground-to-air link is modeled as $l_{\air}(r)=L_o r^{\alpha_{\air}}$  where $\alpha_{\air}$ is its path loss exponent. This assumption facilitates a tractable analysis, and it is further validated for the following 3GPP channel models: RMa-AV for rural areas, UMa-AV for macro urban areas, and UMi-AV for micro urban areas. In particular, let $h$ and  $d_{\operatorname{2D}}$ denote the UAV's altitude and the two-dimensional distance between the device and the UAV, respectively. Then, the LOS probability is given as \cite{3GPP2017d}\footnote{It is assumed here that $22.5<h\leq300$m. We also note that these models also consider the link between the UAV and a base station (BS) at a fixed height. Since devices are generally at lower heights, we add an offset to the UAV's altitude, which is equal to the difference between the BS given height and a device at height 1m.}
\begin{equation}
\mathbb{P}_{\operatorname{LOS}} = \left\{
\begin{array}{ll}
1, & d_{\operatorname{2D}} \leq \gamma_1\\
\frac{\gamma_1}{d_{\operatorname{2D}}} + \exp\left(-\frac{d_{\operatorname{2D}}}{\gamma_2}\right)\left(1-\frac{\gamma_1}{d_{\operatorname{2D}}}\right), &d_{\operatorname{2D}} >\gamma_1
\end{array}\right.,
\end{equation}
where $\gamma_1$ and $\gamma_2$ are constants that depend on the environment \cite[Table B-1]{3GPP2017d}. Fig. \ref{fig:PLOS_vs_height} shows the LOS probability with variations of the UAV's altitude. It is observed that the LOS probability is typically very high, particularly in rural and macro urban areas. Since micro urban areas are typically dense, we can expect that the UAV is at a higher altitude, making the LOS assumption reasonable.

\begin{figure*}[t!]
	\centering
	\begin{subfigure}[t]{.32\textwidth}
		\centering
		\includegraphics[width=2in]{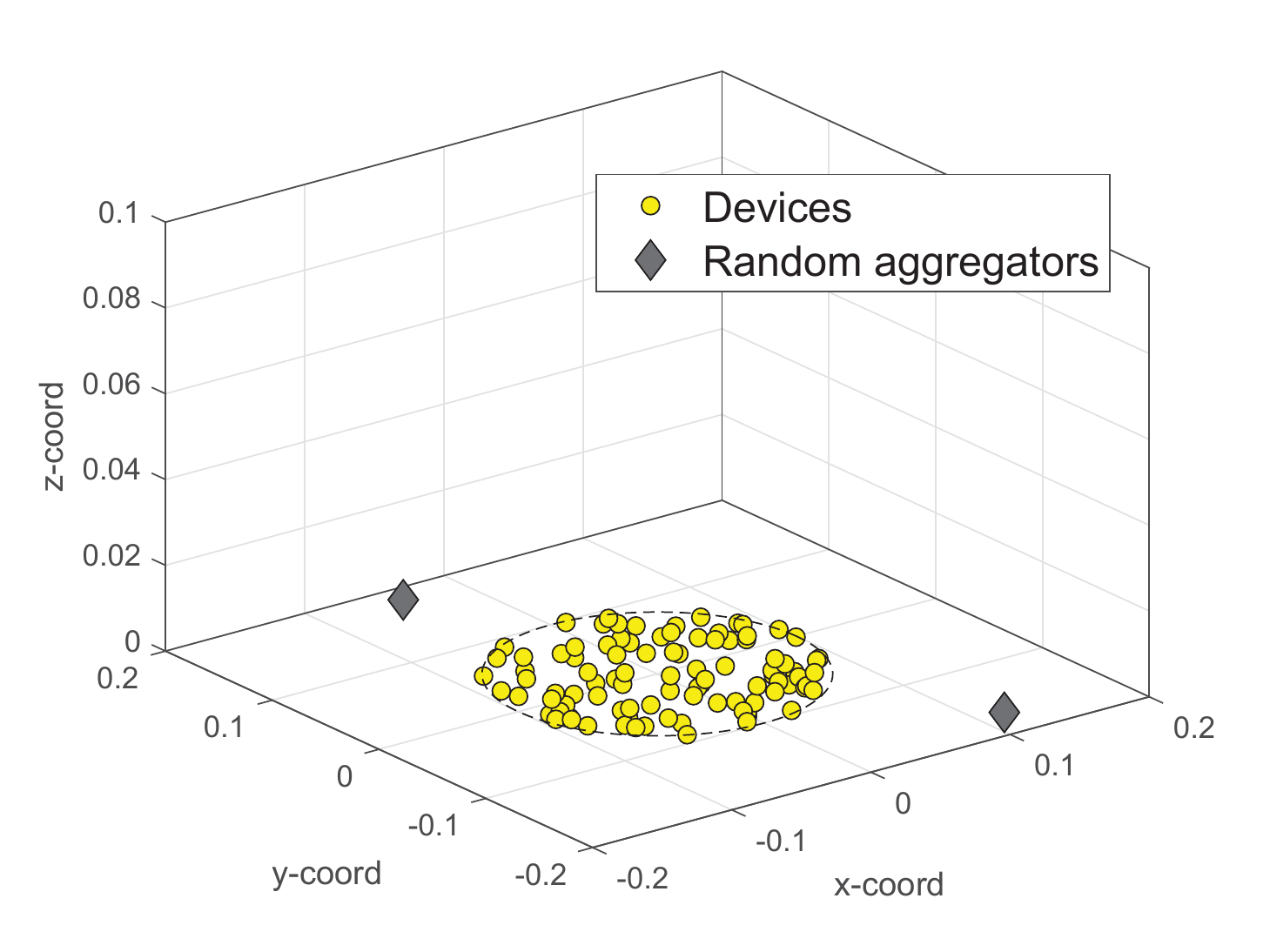}
		\caption{Random deployment}
		\label{fig:RAdeployment}
	\end{subfigure}~
	\begin{subfigure}[t]{.32\textwidth}
		\centering
		\includegraphics[width=2in]{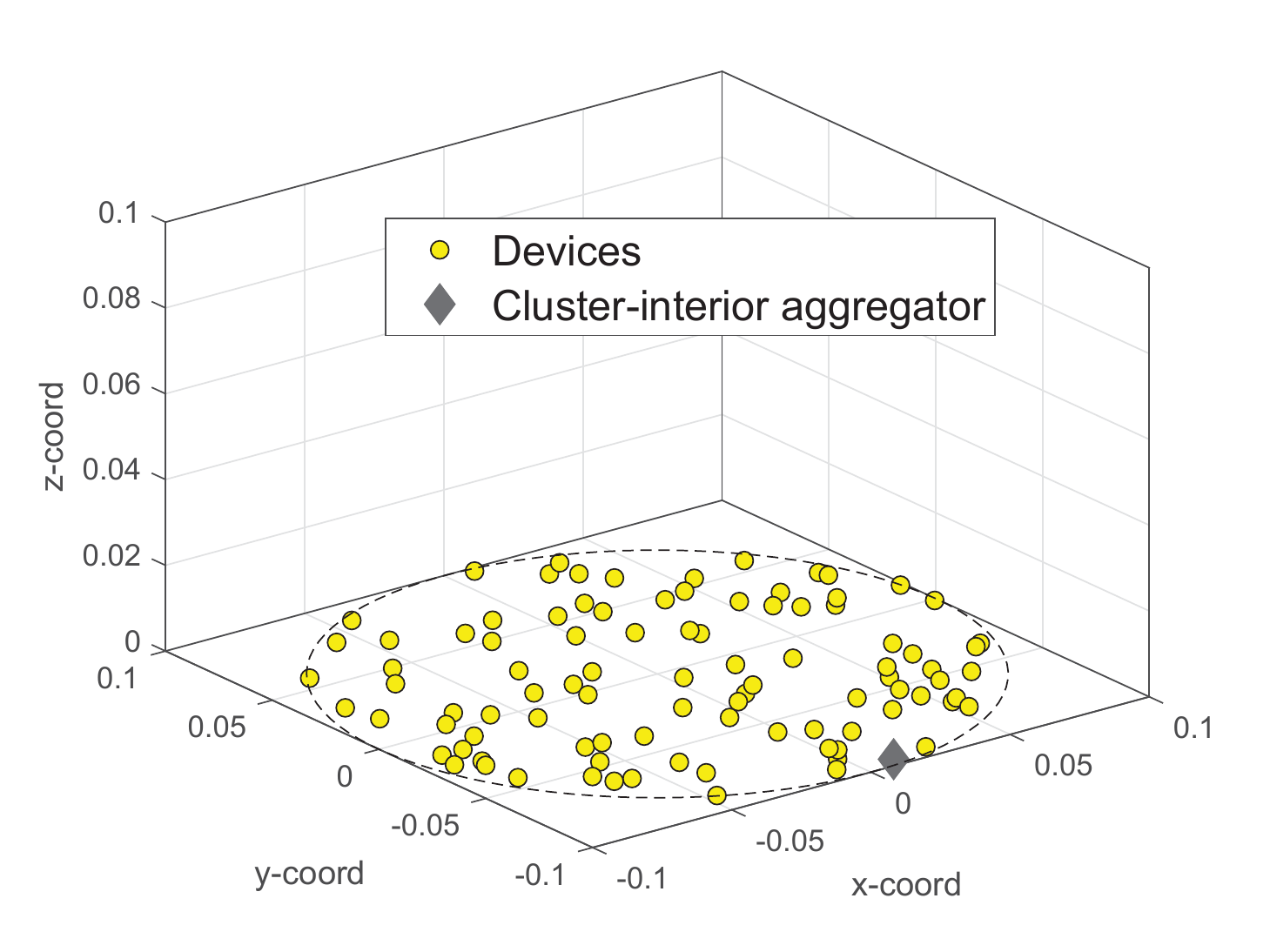}
		\caption{Cluster-interior deployment}
		\label{fig:CIdeployment}
	\end{subfigure}~
	\begin{subfigure}[t]{.32\textwidth}
		\centering
		\includegraphics[width=2in]{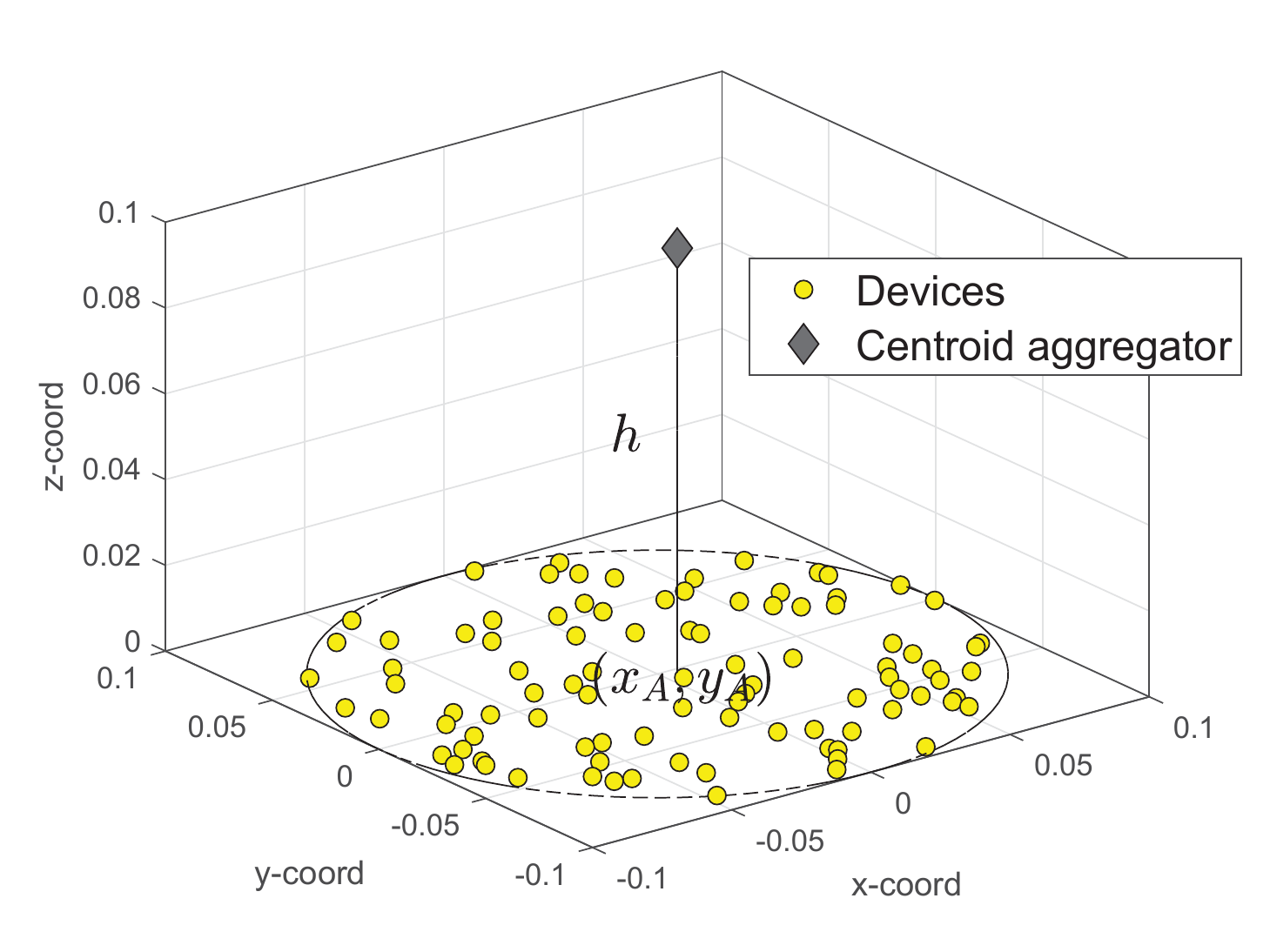}
		\caption{Centroid deployment}
		\label{fig:CNdeployment}
	\end{subfigure}
	\caption{Three different deployment strategies are studied.} 
	\vspace{-.1in}
\end{figure*}	

\subsection{Deployment Strategies} 
We study three different deployment strategies, which are presented next.

\subsubsection{Random deployment}
In this deployment strategy, terrestrial aggregators are randomly deployed in the region without utilizing prior knowledge on the locations of the devices. Hence, we model the locations of these aggregators using the HPPP with density $\lambda$. The device connects to the nearest aggregator. This deployment strategy can be interpreted as a cellular operator densifying a coverage area with data aggregators or utilizing existing  small cells. This deployment is shown in Fig. \ref{fig:RAdeployment}, and it has been widely assumed in the literature, but under different distributions of devices \cite{KwonCioffi2013,MalakAndrews2016}.

\subsubsection{Cluster-interior deployment}
In this deployment strategy, $N$ terrestrial aggregators are randomly deployed inside the same cluster that contains the devices. Each device connects to its nearest aggregator. Such deployment strategy can be interpreted as: (i) randomly assigning a device to act as a cluster head, or (ii) the environment is densely occupied by obstacles that hinder deploying the aggregator at a desired position. Fig.  \ref{fig:CIdeployment} shows the cluster-interior deployment with a single cluster head.

%\footnote{Perhaps a more practical model would be to have the aggregator randomly deployed in a disk of radius $D\ll R$, which is centered around a desired position inside the cluster region. However, such model leads to intractable analysis.}

\subsubsection{Centroid deployment}
In this deployment strategy, the locations of all devices are used to position the aggregator at the point that minimizes the sum of distances squared (between devices and the aggregator), i.e., the $(x_{\operatorname{A}},y_{\operatorname{A}})$ coordinate of the aggregator is the solution of
\begin{equation}
\label{eq:minimization}
(x_{\operatorname{A}},y_{\operatorname{A}}) = \operatorname*{argmin}_{(x,y)}  \mathbb{E}\left[\sum_{i\in\Phi}(x_i-x)^2 + (y_i-y)^2\right],
\end{equation}
where $\Phi$ is the set of devices' locations. It can be shown that the solution to this minimization problem is the centroid of the cluster, i.e., $(x_c,y_c)$, since $\mathbb{E}\left[(x_i-x)^2 + (y_i-y)^2\right]
= \frac{R^2}{2} + (x-x_c)^2+(y-y_c)^2$.
We consider two centroid deployments: a terrestrial one and an aerial one, where the UAV is positioned at $(x_{\operatorname{A}},y_{\operatorname{A}},h)$.\footnote{Since we only consider a LOS component, then the optimal $h$ in terms of the criterion used in (\ref{eq:minimization}) is the smallest $h$ for which the NLOS component remains negligible. Below that, decreasing $h$ further can increase the path loss as the NLOS probability increases.}  Note that the latter deployment is more practical due to the UAV's mobility and the few obstructions incurred at higher altitudes. This deployment strategy is shown in Fig. \ref{fig:CNdeployment}.

\subsection{Performance Metrics}
\subsubsection{Average transmit power consumption} 
At the transmitting device side, we consider the average power consumption as a performance metric. Specifically, we model the power consumption as \cite{MalakAndrews2016}
\begin{equation}
\label{eq:TxPowerConsumption}
P_{\operatorname{TX}} = P_{\operatorname{CP}} + \eta^{-1} P_{T},
\end{equation}
where $P_{\operatorname{CP}}$ denotes the circuitry power consumption, which is assumed to be constant, and $\eta$ denotes the power amplifier efficiency. Hence, different deployment strategies are compared in terms of $\bar P_{\operatorname{TX}} = \mathbb{E}[P_{\operatorname{TX}}]$, where the expectation is with respect to the distance $r$. To determine the lifetime of the device, for a given transmit power consumption, we follow the 3GPP evaluation methodology in \cite{3GPP2015e}. Specifically, in the process of UL transmission, the device operates in four different stages: standby, idle, transmission, and reception. Let $P_{\operatorname{S}}$, $P_{\operatorname{I}}$, and $P_{\operatorname{RX}}$ be the power consumption of the standby, idle, and reception, stages, respectively, and let the duration of each stage be denoted by $T_{\operatorname{S}}$, $T_{\operatorname{I}}$, $T_{\operatorname{RX}}$, and $T_{\operatorname{TX}}$. Let $N_{\operatorname{rep}}$ denote the number of reports per day, then the average total energy consumed per day, in Joules, is 
\begin{equation}
\begin{aligned}
\bar E_{\operatorname{IoT}} &= N_{\operatorname{rep}} \left(T_{\operatorname{TX}}\bar P_{\operatorname{TX}}+T_{\operatorname{RX}}P_{\operatorname{RX}}+T_{\operatorname{I}}P_{\operatorname{I}}\right)+T_{\operatorname{S}}P_{\operatorname{S}}.
\end{aligned}
\end{equation} 
Thus, for a battery capacity of $C_{\operatorname{IoT}}$, given in Wh, the device lifetime, in years, is given as
\begin{equation}
Y = \frac{C_{\operatorname{IoT}}}{\bar E_{\operatorname{IoT}}} \times \frac{3600}{365}.
\end{equation}

\subsubsection{Coupling loss} 
At the aggregator side, we consider the coupling loss as a performance metric, which quantifies the link budget needed for a target signal-to-noise ratio (SNR) and the coverage enhancement needed to arrive at a desired maximum coupling loss (MCL) \cite{CATM2017}. In particular, the MCL is defined as \cite{CATM2017}
\begin{equation}
\label{eq:MCL}
\operatorname{MCL}_{\operatorname{dB}}= S - Q + G,
\end{equation} 
where $S$ is the transmitter's maximum power (dBm), i.e, in this case $S=10\log_{10}(P_{\operatorname{max}})$, $Q$ is the receiver sensitivity (dBm), and $G$ is the gain achieved, in dB,  using some coverage enhancement techniques, e.g., signal repetition or power spectral density boosting \cite{CATM2017}.  The receiver sensitivity is calculated as
\begin{equation}
Q =  N_0 + P_{\operatorname{NF}} + 10\log_{10}(W) + \tau,
\end{equation}
where $N_0$ is the thermal noise density (dBm/Hz), $P_{NF}$ is the noise figure (dB), $W$ is the occupied channel bandwidth (Hz), and $\tau$ is the required SNR (dB), e.g., $\tau=-7.8$dB and $\tau=-4.3$dB for the physical uplink control channel (PUCCH) and shared channel (PUSCH) in cellular networks, respectively \cite{CATM2017}. 

Let $\mu=10^{\operatorname{MCL_{\operatorname{dB}}}/10}$ be the MCL in linear scale, then we define the \emph{IoT coverage} probability as 
\begin{equation}
\label{eq:MCL1}
\mathbb{C}(\mu)= \mathbb{P}(l(r)\leq \mu|S).
\end{equation}  
In other words, the device is considered in coverage if the device-aggregator link path loss is less than $\mu$.

%---------------------------------------------------------------------------------
%                         III. {Analysis of Different Deployment Strategies
%---------------------------------------------------------------------------------
\section{Analysis of Different Deployment Strategies}\label{sec:analysis}
The probability density function (pdf) of the random distance between a typical device and the data aggregator, i.e., $f_{\mathcal{R}}(r)$, is critical in performance evaluation. Indeed, the average transmit power can be rewritten as
\begin{equation}
\begin{aligned}
\bar P_{\operatorname{TX}} 	
							&= P_{\operatorname{CP}} + \eta^{-1} \hat P_o \mathbb{E}[r^{\epsilon\alpha}\mathbf{1}(r\leq \zeta)] +  \eta^{-1}  P_{\operatorname{max}} \mathbb{P}(r> \zeta),
\end{aligned}
\end{equation}
where $\hat P_o = P_o (L_0)^{\epsilon\alpha}$ and $\zeta$ is a threshold, denoted henceforth as \emph{the critical distance}, and it depends on the deployment strategy. Similarly, the IoT coverage can be expressed as 
\begin{equation}
\label{eq:MCL2}
\mathbb{C}(\mu) = \mathbb{P}(r\leq \hat \mu^{1/\alpha}|S),
\end{equation}
where $\hat \mu = \mu/L_o$. In the next sections, we analytically compare the different strategies in terms of the aforementioned metrics. 

\subsection{Random deployment}
In this strategy, the pdf of the distance between the terrestrial aggregator and a typical device can be shown to be as follows. 
\begin{lemma} 
	The pdf of the distance between a device and the nearest aggregator is given by
	\begin{equation}
	\label{eq:RAdistancePDF}
	f_{\operatorname{RA}}(r)= 2\pi \lambda r \exp\left(-\pi\lambda r^2\right).
	\end{equation}
\end{lemma}
\noindent This follows using the void probability of the HPPP \cite{JoAndrews2012}. Using (\ref{eq:RAdistancePDF}), the performance of the IoT device under this deployment strategy is given as follows.

\begin{theorem}
	The average transmit power consumed in the UL under random deployment is given by
	\begin{equation}
	\begin{array}{ll}
	\label{eq:avgPowerRA}
	\bar P_{\operatorname{TX}}^{\operatorname{RA}}&=
	 P_{\operatorname{CP}} + \eta^{-1} \hat P_o \left(\pi\lambda\right)^{-\frac{\epsilon\alpha_{\g}}{2}}\gamma\left(\frac{q_2}{2},\pi\lambda \zeta_{\operatorname{RA}}^2\right)\\
	 &+\eta^{-1} P_{\operatorname{max}} \exp(-\pi\lambda \zeta_{\operatorname{RA}}^2),
	\end{array}
	\end{equation}
	where  $\zeta_{\operatorname{RA}}=(\frac{P_{\operatorname{max}}}{\hat P_o})^{\frac{1}{\epsilon\alpha_{\g}}}$, $q_i = i + \epsilon\alpha_{\g}$, and $\gamma(\cdot,\cdot)$ is the lower incomplete Gamma function. Further, the IoT coverage is given as
	\begin{equation}	
	\label{eq:MCL_RA}
	\mathbb{C}_{\operatorname{RA}}(\mu) = 1 - \exp(-\pi \lambda\hat \mu^{2/\alpha_{\g}}).
	\end{equation}
\end{theorem}
\noindent Proof: See the Appendix. \hfill $\blacksquare$ 

\noindent \emph{Remarks:} When $P_{\operatorname{max}}\rightarrow\infty$,  we can simplify (\ref{eq:avgPowerRA}) as follows 
\begin{equation}
\bar P_{\operatorname{TX},\infty}^{\operatorname{RA}}=P_{\operatorname{CP}} + \eta^{-1} \hat P_o \left(\pi\lambda\right)^{-\frac{\epsilon\alpha_{\g}}{2}}\Gamma\left(\frac{q_2}{2}\right).
\end{equation} 
The key insight here is that power consumption under this strategy is proportional to $\lambda^{-\epsilon}$. In addition, the minimum density of data aggregators required to achieve an MCL of $\mu$ with a probability of $\beta$ is 
\begin{equation}
\lambda^\star  = \frac{\ln\left(\frac{1}{1-\beta}\right)}{\pi  \mu^{2/\alpha_{\g}}}. 
\end{equation}
Clearly, $\lambda^\star$ increases with the path loss exponent, and hence urban environments require higher density of data aggregators. 

\subsection{Cluster-interior deployment}
In this strategy, the pdf of the distance between the aggregator and a typical device is equivalent to the distribution of the distance between two randomly deployed points in a circle of radius $R$, which is given in the following lemma.

\begin{lemma} 
	The pdf of the distance between a device and a randomly deployed aggregator inside a disk of radius $R$ is given by \cite{Garcia2005}
	\begin{equation}
	\label{eq:CIdistancePDF}
	f_{\operatorname{CI}}(r)= 
	\left\{\begin{array}{ll}
	\frac{4r}{\pi R^2} \cos^{-1}\left(\frac{r}{2R}\right) - \frac{2r^2}{\pi R^4} \sqrt{R^2-\frac{1}{4}r^2}, &r\leq 2R\\
	0																									  , &r>2R
	\end{array}\right.																						
	\end{equation}
\end{lemma}

Using (\ref{eq:CIdistancePDF}), we can obtain the following expressions of power consumption and IoT coverage. 

\begin{theorem}
The average transmit power consumed in the UL under cluster-interior deployment with $N$ aggregators is 
\begin{equation}
\label{eq:avgPowerCI}
\begin{array}{ll}
\bar P_{\operatorname{TX}}^{\CI}&= P_{\operatorname{CP}} + \eta^{-1}  N \Bigg(P_{\operatorname{max}} \displaystyle\int_{\zeta_{\CI}}^{2R} f_{\CI}(r)\left(1-\Psi(r)\right)^{N-1} dr\\ 
&+\hat P_o \displaystyle\int_0^{\zeta_{\CI}} r^{\epsilon\alpha_{\g}}f_{\CI}(r)\left(1-\Psi(r)\right)^{N-1} dr\Bigg),
\end{array}
\end{equation}
where  $\zeta_{\CI}=\min\{2R,(\frac{P_{\operatorname{max}}}{\hat P_o})^{\frac{1}{\epsilon\alpha_{\g}}}\}$ and 
\begin{equation}
\begin{aligned}
\Psi(r) &= \frac{1}{\pi}\Big(\textstyle 4 \csc^{-1}\left(\frac{2R}{r}\right) + 2\left(\frac{r}{R}\right)^2\sec^{-1}\left(\frac{2R}{r}\right)\\
		&-\textstyle  2 \tan^{-1}\left(\frac{r}{\sqrt{4R^2-r^2}}\right) - \frac{r(r^2+2R^2)\sqrt{4R^2-r^2}}{4R^4}\Big).
\end{aligned}
\end{equation}
The IoT coverage probability is given as
\begin{equation}
\begin{aligned}
\label{eq:MCLCI}
\mathbb{C}_{\CI}(\mu) &= N \int_{0}^{\min\{\mu^{\frac{1}{\alpha_{\g}}},2R\}} f_{\CI}(r)\left(1-\Psi(r)\right)^{N-1} dr\\
&\stackrel{(N=1)}{=} \min\left\{\Psi(\min\{\mu^{\frac{1}{\alpha_{\g}}},2R\}),1\right\}.
\end{aligned}
\end{equation}
\end{theorem}
\noindent Proof: See the Appendix. \hfill $\blacksquare$ 

\noindent\emph{Remarks:} Assuming $N=1$ and unbounded transmit power, then we can simplify (\ref{eq:avgPowerCI}) to 
%\begin{equation}
%\label{eq:avgPowerCIOne}
%\begin{array}{ll}
%\bar P_{\operatorname{TX}}^{\CI}= P_{\operatorname{CP}} + \eta^{-1} \hat P_o \frac{\zeta_{\CI}^{q_2}}{\pi R^3 q_2q_3}\bigg[q_3\cdot R \left(\pi -\csc^{-1}\left(\frac{2R}{\zeta_{\CI}}\right)\right)\\
%-\zeta_{\CI}\left(q_2 \tilde F(-\frac{1}{2}) -\tilde F(\frac{1}{2})\right)\bigg]+\eta^{-1}P_{\operatorname{max}} (1-\Psi(\zeta_{\CI})),
%\end{array}
%\end{equation}
%where  , $\tilde F(x)={}_2F_1(x,\frac{1}{2}q_3;\frac{1}{2}q_5;\frac{\zeta_{\CI}^2}{R^2})$ is the Gauss hypergeoemtric function. Furthermore, for $P_{\operatorname{max}}\rightarrow\infty$, we have
\begin{equation}
\label{eq:avgPowerCIinf}
\bar P_{\operatorname{TX},\infty}^{\CI} =  P_{\operatorname{CP}} + \eta^{-1} \hat P_o \left(\frac{8\Gamma(\frac{q_3}{2})}{\sqrt{\pi}q_2 \Gamma(\frac{q_6}{2})}\right)(2R)^{\epsilon\alpha_{\g}}.
\end{equation}
The insight here is that the average power is proportional to $(2R)^\epsilon$. Hence, for more dispersed devices over space, using one cluster-interior aggregator becomes inefficient.	

\subsection{Centroid deployment}
Recall that the aerial aggregator is located at the centroid, and thus we can obtain the pdf of the distance between the device and the aggregator using a pdf transformation of the  distance between the center of a disk and a random point on that disk. This is shown in the following lemma. 

\begin{lemma} 
	The pdf of the distance between a device inside a disk of radius $R$  and an aerial aggregator at altitude $h$ is given by
	\begin{equation}
	\label{eq:CNdistancePDF}
	f_{\operatorname{CN}}(r)= 
	\left\{\begin{array}{ll}
	\frac{2r}{R^2}, &h\leq r\leq \sqrt{R^2+h^2}\\
	0,				&\text{otherwise}.
	\end{array}\right.																						
	\end{equation}
\end{lemma}
Using (\ref{eq:CNdistancePDF}), the performance of the device under the centroid deployment is given as follows. 

\begin{theorem}
	The average transmit power consumed in the UL under the centroid deployment is given by
	\begin{equation}
	\label{eq:avgPowerCN}
	\begin{aligned}
	\bar P_{\operatorname{TX}}^{\CN}&= P_{\operatorname{CP}} + \eta^{-1}\hat P_o   \frac{2}{p_2 R^2} \left(\zeta_{\CN}^{p_2}-h^{p_2}\right)\\
	&+\eta^{-1} P_{\operatorname{max}}\frac{R^2+h^2-\zeta_{\CN}^2}{R^2},
	\end{aligned}
	\end{equation}
	where  $\zeta_{\CN}=\min\{\sqrt{R^2+h^2},\max\{h,(\frac{P_{\operatorname{max}}}{\hat P_o})^{\frac{1}{\epsilon\alpha_{\air}}}\}\}$ and $p_i=i+\epsilon \alpha_{\air}$. In addition, the IoT coverage is given as
	\begin{equation}
	\begin{aligned}
	\label{eq:avgMCLCN}
	\mathbb{C}_{\CN}(\mu) &= \frac{\min\{\hat \mu^{2/\alpha_{\air}},R^2+h^2\}-\min\{\hat \mu^{2/\alpha_{\air}},h^2\}}{R^2}.
	\end{aligned}
	\end{equation}
\end{theorem}	
\noindent Proof: See the Appendix. \hfill $\blacksquare$

\noindent \emph{Remarks:} The performance of a terrestrial aggregator can be found by making $h\rightarrow0$ and replacing $\alpha_{\air}$ with $\alpha_{\g}$. In addition, the average power consumption in (\ref{eq:avgPowerCN}) under the unbounded transmit power is given as
\begin{equation}
\label{eq:avgPowerCNinfA}
\begin{aligned}
\bar P_{\operatorname{TX},\infty}^{\CN}&= P_{\operatorname{CP}} + \eta^{-1}\hat P_o   \frac{2}{p_2 R^2} \left((R^2+h^2)^{p_2/2}-h^{p_2}\right).
\end{aligned}
\end{equation}
Thus, comparing (\ref{eq:avgPowerCNinfA}) with (\ref{eq:avgPowerCIinf}), we observe that centroid deployment of the aggregator scales with $R^{\epsilon}$ compared to the $(2R)^\epsilon$ scaling when the aggregator is randomly deployed inside the cluster. 
%As for the IoT coverage in (\ref{eq:avgMCLCN}), we typically have $\mu_{\CN}^{2/\alpha_{\air}}\gg R^2+h^2$, and hence $\mathbb{C}_{\CN}(\mu)\rightarrow1$, showing that aerial aggregators provide superior coverage, given that the LOS component is dominant.  

%Table \ref{tab:comparison} shows a comparison among the different deployment schemes in terms of: (i) the transmit power consumption saving relative to that achieved using cluster-interior deployment and (ii) the  impact of doubling the radius $R$ on the average power consumption. Here, it is assumed that there is no transmit power bound, $\epsilon=1$, $h=R$, $\alpha_{\g}=4$, $\alpha_{\air}=2$, and $N=1$. 
%
%
%
%
%
%
%\begin{table}[!t]
%	\caption{Comparison of different schemes}
%	\label{tab:comparison}
%	\centering
%	\footnotesize
%	\begin{tabular}{|>{\centering\arraybackslash}m{0.5in}|>{\centering\arraybackslash}m{0.5in}|>{\centering\arraybackslash}m{0.5in}|>{\centering\arraybackslash}m{0.5in}|>{\centering\arraybackslash}m{0.5in}|}
%		\hline
%		\rowcolor{lightgray}Comparison &Cluster-interior & Centroid (terrestrial)  	&Centroid (aerial) &Random \\\hline
%		Relative saving (dB)           &0     & $6.5$                     &$20\log(R)$ &Depends on $\lambda$\\\hline
%		Doubling $R$				   &16x	  & 16x					    &6x             &Independent		\\\hline
%	\end{tabular}
%\end{table}
%%

%---------------------------------------------------------------------------------
%                         IV. Simulation Results
%---------------------------------------------------------------------------------
\section{Simulation Results}\label{sec:sim}
We validate the theoretical results via Monte Carlo simulations, where we run 1000 realizations, each with different locations of devices. We use lines and markers to denote the theoretical expressions and the simulations, respectively. A summary of the simulation parameters is given in Table \ref{tab:parameters}. It is assumed that each device transmits a packet of size 200bytes every two hours \cite{3GPP2015e}, and the durations are based on the UL specifications when the MCL is 154dB \cite{3GPP2015e}.

\begin{table}[!t]
	\caption{Main parameters}
	\label{tab:parameters}
	\centering
	\begin{tabular}{|c|l|}
		\hline
		\textbf{Description }  				&  \textbf{Parameters}\\\hline
		Path loss 					& $\alpha_{\g}=3.5$, $\alpha_{\air}=2.2$ \cite{3GPP2017d} , and $L_{0,\operatorname{dB}}\approx38$dB \\\hline
		UAV altitude			    & $h=100$m	\\	\hline
		Tx power model 				& $P_{\operatorname{CP}}=90$mW and $\eta=0.44$ \cite{3GPP2015e}\\\hline
		\multirow{3}{*}{IoT device} & NB-IoT with $W=180$KHz,  $P_{\operatorname{max,dBm}}=20$dBm,\\	
		& $P_{o,\operatorname{dBm}}=-46$dBm (or  $-100$dBm/Hz)\\
		&$C_{\operatorname{IoT}}=5$Wh\cite{3GPP2015a}, and $N_{\operatorname{rep}}=12$\\\hline
		\multirow{2}{*}{Powers}     & $P_{\operatorname{RX}}=90$mW, $P_{\operatorname{I}}=3$mW, \\
		&and $P_{\operatorname{S}}=0.015$mW \cite[Table 1]{3GPP2015e}\\\hline
		\multirow{2}{*}{Durations} 	& $T_{\operatorname{TX}}=983$ms, $T_{\operatorname{RX}}=565$ms, \\
		&$T_{\operatorname{I}}=22451$ms, and $T_{\operatorname{S}}=86400$s 	\cite[Table 6]{3GPP2015e}		\\	 	\hline	
		\multirow{2}{*}{MCL}	    & $N_0=-174$dBm/Hz, $\tau=-4.3$dB, $P_{\operatorname{NF}}=5$,\\
		&and $\operatorname{MCL}_{\operatorname{dB}}=154$dB \cite{CATM2017}\\\hline
	\end{tabular}
\end{table}

\begin{figure}[t!]
	\centering
	\begin{subfigure}[t]{.5\textwidth}
		\centering
		\includegraphics[width=3.25in]{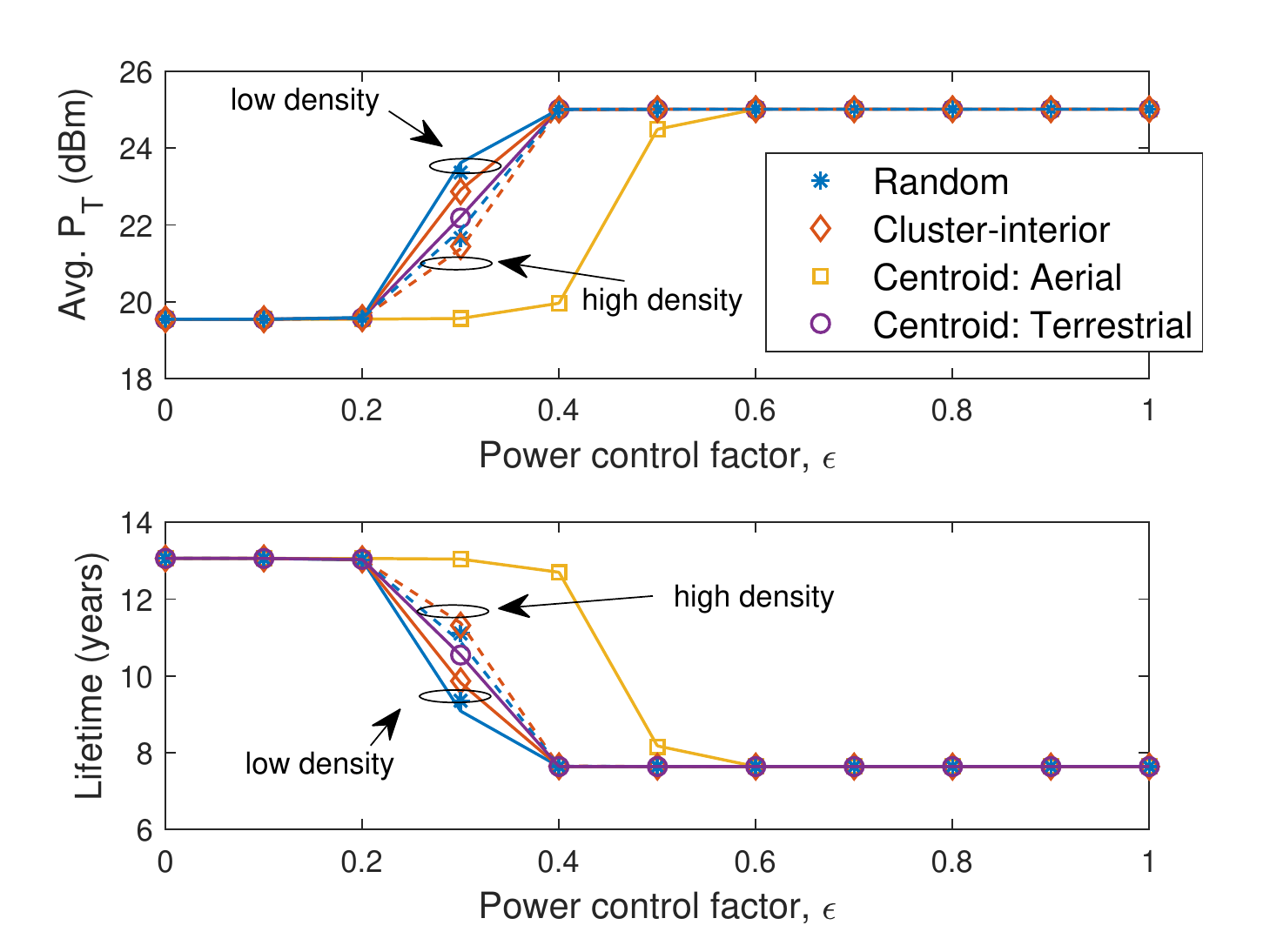}
		\caption{Impact of the PCF $\epsilon$ ($R=200$m)}
		\label{fig:performance_vs_PCF}
	\end{subfigure}\\
	\begin{subfigure}[t]{.5\textwidth}
		\centering
		\includegraphics[width=3.25in]{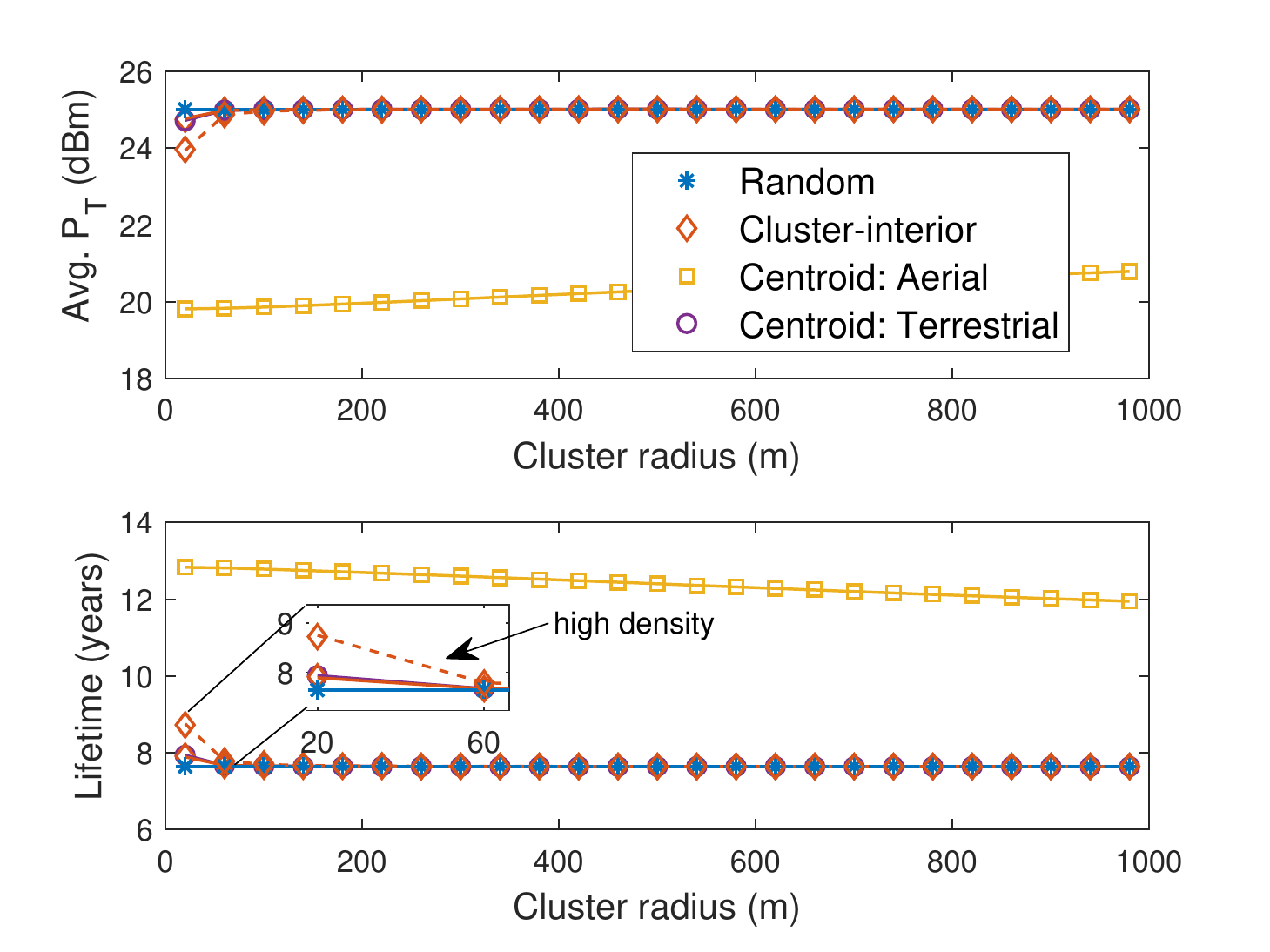}
		\caption{Impact of cluster radius $R$ ($\epsilon=0.4$)}
		\label{fig:performance_vs_radius}
	\end{subfigure}
	\caption{Comparison of different deployment strategies in terms of transmit power consumption and device lifetime.} 
	\vspace{-0.05in}
\end{figure}

We first study the average power consumption and device lifetime with variations of the PCF, as shown in Fig. \ref{fig:performance_vs_PCF}. We vary the PCF because, in general, this parameter is set by the network for each device \cite{3GPP2015a}, and thus we want to study the performance for the different possible values.  We consider two densities for random and cluster-interior strategies. In the low density, we assume $N=1$ and $\lambda=5$/km$^2$, whereas in the high density, we consider $N=5$ and $\lambda=25$/km$^2$. It is evident that the theoretical curves match well with simulations. In addition, centroid deployment of a single terrestrial aggregator outperforms the cluster-interior deployment with a single cluster head, showing that optimizing the location of the aggregator tangibly increases the device lifetime. The gains of centroid deployment are more significant when using aerial aggregators instead of terrestrial ones as ground-to-air links have higher LOS and low path loss exponent \cite{3GPP2015a}. To improve the performance of random and cluster-interior strategies, a higher density of aggregators is needed. We remark that all schemes have the same performance when $\epsilon$ is very large, as a higher value implies higher compensation of the path loss, forcing the device to transmit at $P_{\operatorname{max}}$, which is constant for all schemes. Similar trends follow when $\epsilon$ is very small, as the transmit power becomes dominated by $P_{o}$ when $\epsilon\rightarrow0$.

Fig. \ref{fig:performance_vs_radius} shows the performance for different cluster sizes. Recall that random deployment of aggregators is independent of the cluster radius (cf. (\ref{eq:avgPowerRA})). It is observed that the centroid deployment of terrestrial aggregators is useful when devices are more clustered. Since increasing the radius $R$ makes devices appear randomly deployed over space, instead of being clustered, the centroid terrestrial deployment performs very similarly to the cluster-interior deployment. Last, aerial aggregators are shown to be more robust to different cluster sizes due to the low path loss of ground-to-air links. 

\begin{figure}[t!]
	\centering
	\begin{subfigure}[t]{.5\textwidth}
		\centering
		\includegraphics[width=3.2in]{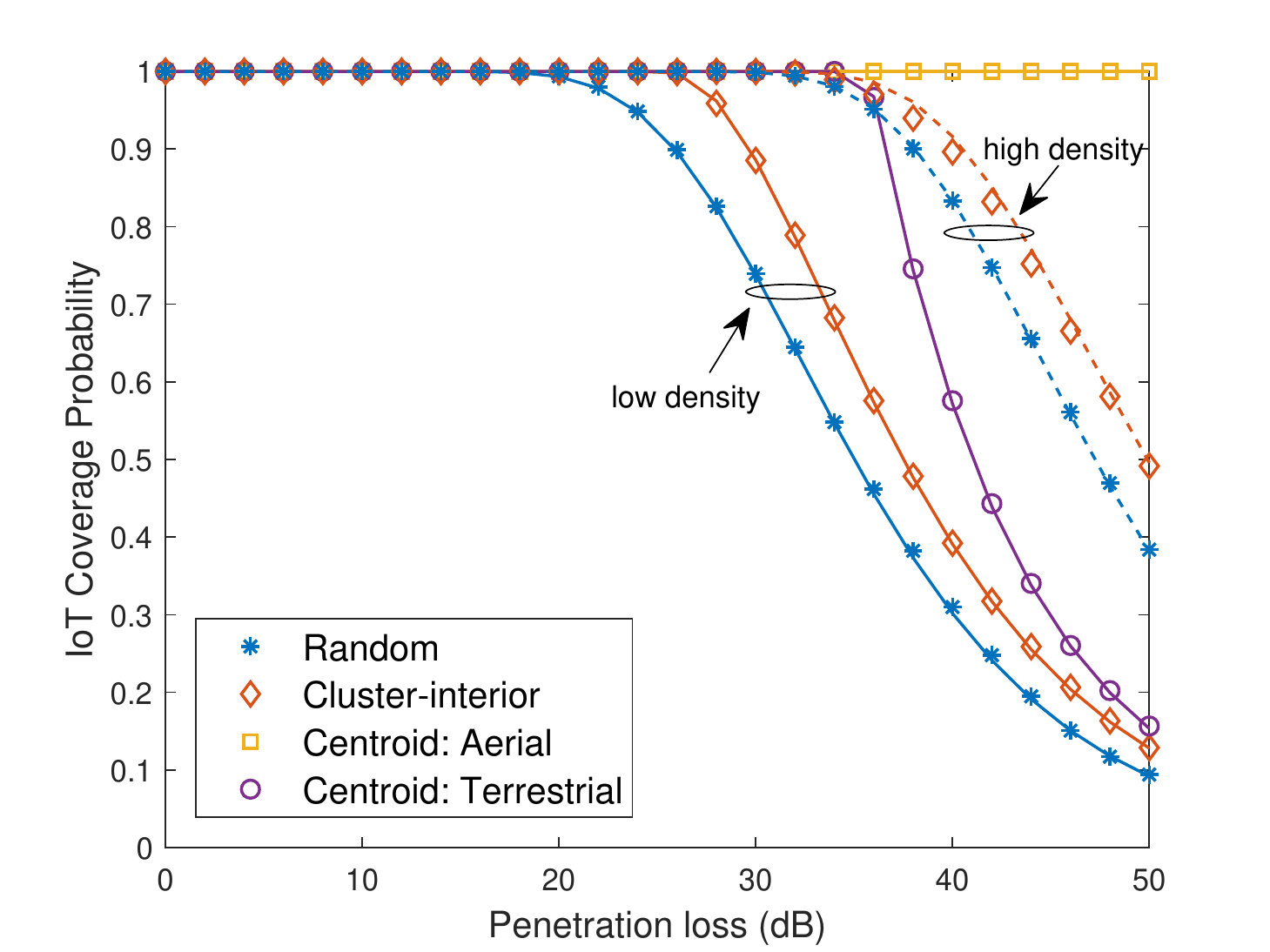}
		\caption{Variations of penetration losses ($R=200$m)}
		\label{fig:MCL_vs_PenLoss}
	\end{subfigure}\\
	\begin{subfigure}[t]{.5\textwidth}
		\centering
		\includegraphics[width=3.2in]{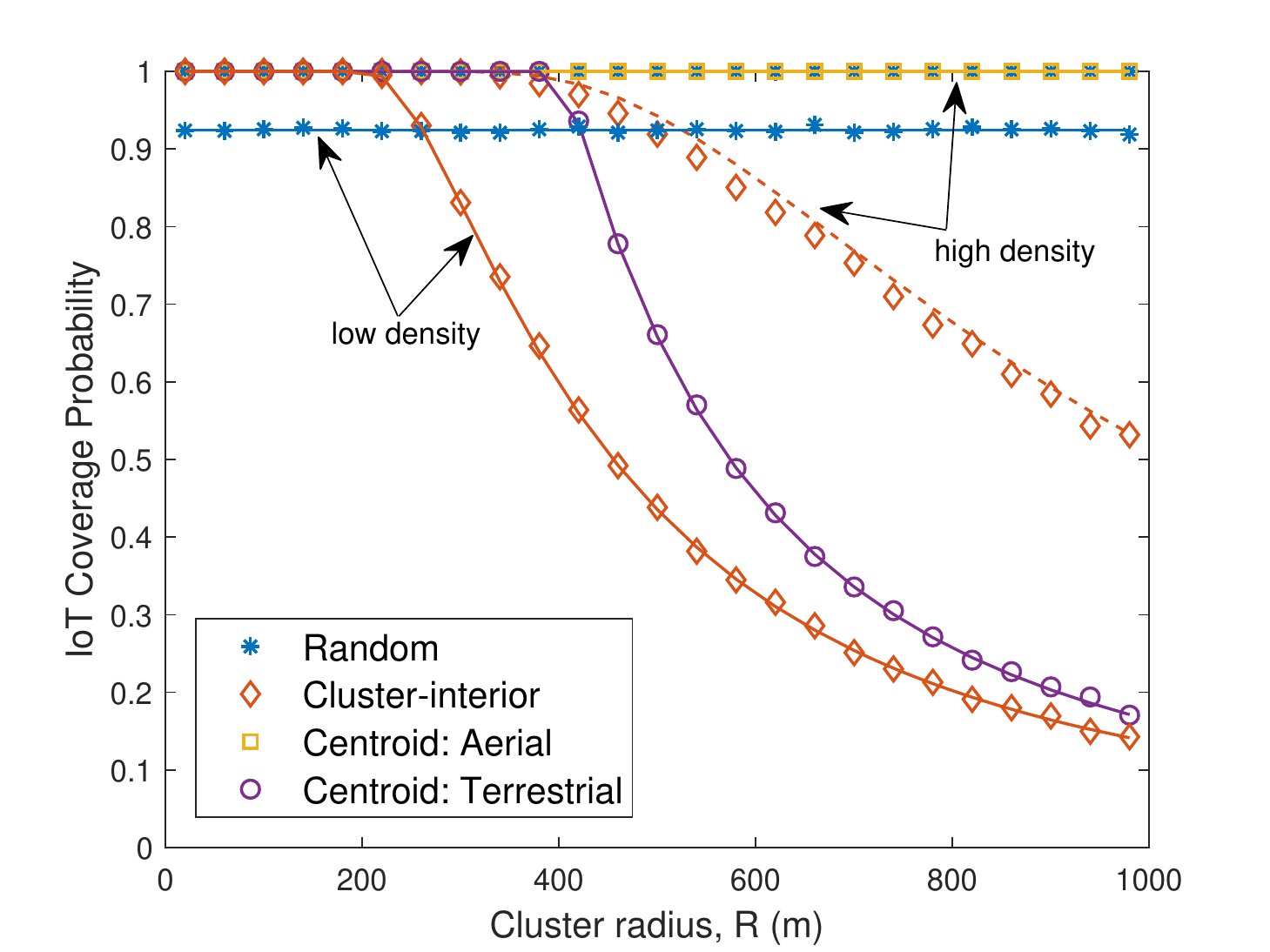}
		\caption{Variations of $R$ (penetration losses are fixed at 25dB)}
		\label{fig:MCL_vs_radius}
	\end{subfigure}
	\caption{IoT coverage performance comparison.} 
	\vspace{-0.1in}
\end{figure}

Fig. \ref{fig:MCL_vs_PenLoss} shows the IoT coverage probability in the presence of different penetration losses, where the target maximum coupling loss is $\operatorname{MCL}_{\operatorname{dB}}=154$dB. It is shown that densification of randomly deployed aggregators is necessary to support deep coverage. It is also observed that aerial aggregators can support very deep coverage requirements thanks to its proximity to devices and the low ground-to-air propagation losses. In Fig. \ref{fig:MCL_vs_radius}, we study the IoT coverage for different cluster radii. Random deployment of aggregators can provide high coverage when they are densely deployed over space, irrespective whether devices are clustered or not. In contrary, the terrestrial centroid deployment performs poorly once devices become dispersed over space, i.e., randomly deploying multiple aggregators is more coverage-efficient than deploying a single aggregator with optimized location.

%---------------------------------------------------------------------------------
%                         V. Conclusion and Future Work
%---------------------------------------------------------------------------------
\section{Conclusion}\label{sec:conclusion}
Data aggregation is an attractive solution to support cellular IoT communications as it helps in improving coverage and extending device lifetime. A critical aspect to data aggregation is how to deploy aggregators over space. In this paper, we use stochastic geometry to study and analyze three different deployment strategies in terms of cellular IoT-specific metrics, following the same 3GPP evaluation methodology. In particular, we derive closed-form theoretical expressions of the average transmit power consumption and the IoT coverage. It is shown that the coverage achieved with aerial aggregators is superior to that achieved with terrestrial ones. However, several implementation issues may hinder practical implementation of aerial aggregators, e.g., the drone typically has a short lifetime over the air, its trajectory must be optimized, etc. For these reasons, terrestrial aggregators are still useful due to their ease of implementation. In case devices are clustered, a single, yet location-optimized, terrestrial aggregator is sufficient. As devices become more dispersed  over space, it becomes more beneficial to randomly deploy aggregators at a high density.

\appendix{
\subsubsection{Random deployment}
We have from (\ref{eq:TxPowerConsumption}), 
\begin{equation}
\begin{aligned}
\mathbb{E}\left[r^{\epsilon\alpha_{\g}}\right]&=2\pi\lambda \int_{0}^{\zeta_{\operatorname{RA}}} r^{1+\epsilon \alpha_{\air}}\exp(-\pi\lambda r^2)dr.
\end{aligned}
\end{equation}
Using $\int x^m \exp(\beta x^n)dx = -\frac{\Gamma(\gamma,\beta x^n)}{n\beta^\gamma}$, where $\gamma=\frac{m+1}{n}$ and $2\pi\lambda \int_{\zeta_{\operatorname{RA}}}^{\infty} r\exp(-\pi\lambda r^2)dr=\exp(-\pi \lambda \zeta_{\operatorname{RA}}^2)$  we arrive at (\ref{eq:avgPowerRA}). Similar approach is used to derive the IoT coverage and the power consumption with unbounded transmit power. 

\subsubsection{Cluster-interior deployment}
Let $R_i$ be the distance between a device and the $i$-th aggregator, and $F(n)$ be the cumulative distribution function (CDF). Then, the distribution of the distance between the device and the nearest aggregator is given as $F_N(r)  = \mathbb{P} \left(\operatorname*{min}_i R_i \leq r\right)= 1-[1-F(n)]^N$, and thus the pdf is given as $f_N(r)= N f_{\operatorname{CI}} (r) [1-F(r)]^{N-1}$. Since $F(r)=\int_{0}^r f_{\operatorname{CI}}(r) dr$, we can be show $F(r)= \Psi(r)$. Since $\mathbb{C}_{\operatorname{CI}}(\mu)$ is the CDF of $f_N(r)$, we arrive at (\ref{eq:MCLCI}).

\subsubsection{Centroid deployment}
Note in (\ref{eq:CNdistancePDF}) that the distance is bounded between $h$ and $\sqrt{R^2+h^2}$. Thus, the cutoff distance $\zeta_{\CN}\leq\sqrt{R^2+h^2}$. The lower bound on the cutoff distance similarly follows. Hence, $\mathbb{E}\left[r^{\epsilon\alpha_{\g}}\right]=\frac{2}{R^2} \int_{h}^{\zeta_{\CN}} r^{1+\epsilon\alpha_{\air}}dr$ can be directly evaluated to get (\ref{eq:avgPowerCN}). The derivations of the IoT coverage follows similarly.

%---------------------------------------------------------------------------------
%                         References
%---------------------------------------------------------------------------------
\bibliographystyle{IEEEtran}
\bibliography{C:/Users/ghait/Dropbox/References/IEEEabrv,C:/Users/ghait/Dropbox/References/References}

\end{document}